\pdfoutput=1
\documentclass[11pt]{article}
\usepackage[utf8]{inputenc}  
\usepackage[T1]{fontenc}     
\usepackage[british]{babel}  
\usepackage[colorlinks=true]{hyperref}
\hypersetup{colorlinks=true,linkcolor=teal,citecolor=blue,urlcolor=blue}
\usepackage[margin=10pt,font=small,format=hang,figurename=Fig. ,indention=-.55cm,labelfont={bf,sf,sc}]{caption}
\usepackage{graphicx}
\usepackage{amsmath}
\usepackage{amssymb}
\usepackage{hyperref}
\usepackage{latexsym}
\usepackage{color}
\usepackage{subfigure}

\usepackage{pst-all}
\usepackage{ifpdf}
\usepackage{multimedia}
\usepackage{lmodern}
\usepackage{cite}
\usepackage{dsfont}
\usepackage{bbm}
\usepackage{lipsum}
\usepackage{url}
\usepackage{textcomp}
\usepackage{bbold}
\usepackage{mathtools}
\usepackage{nccmath}
 \usepackage[normalem]{ulem}
\usepackage{ifpdf}
\ifpdf
\else
\fi
  \hypersetup{pdfstartview=XYZ}
\newcommand{\bra}{\begin{array}}
\newcommand{\era}{\end{array}}
\newcommand{\beq}{\begin{equation}}
\newcommand{\eeq}{\end{equation}}
\newcommand{\beqar}{\begin{eqnarray}}
\newcommand{\eeqar}{\end{eqnarray}}

\def\BC{\bb C}
\def\_\BC{\bbi C}

\def\( {\left(}
\def\) {\right)}
\def\[ {\left[}
\def\] {\right]}
\def\no2 {{\textstyle{n\over 2}}}

\newcommand{\lb}{\label}
\begin{document}
\thispagestyle{empty}
\begin{center}

\vspace{1.8cm}

 {\Large {\bf Quantum Fisher information and skew information correlations in dipolar spin system }}\\

\vspace{1.5cm} {\bf R. Muthuganesan}{\footnote {
email: {\sf rajendramuthu@gmail.com}}} and {\bf V. K. Chandrasekar}{\footnote { email: {\sf
chandru25nld@gmail.com}}} \\
\vspace{0.5cm}
{\it Center for Nonlinear Science \& Engineering, School of Electrical \& Electronics Engineering,
SASTRA Deemed University, Thanjavur, Tamil Nadu 613 401, India}\\ [0.5em]

\end{center}
\baselineskip=18pt
\medskip
\vspace{3cm}
\begin{abstract}
Quantum Fisher information (QFI) and skew information (SI) plays a key role in the quantum resource theory. Understanding these measures in the physical system has practical significance in the state parameter estimation and quantum metrology. In this article, we consider a pair of spin-1/2 particles coupled with dipolar and Dzyaloshinsky-Moriya (DM) interactions, serving as the physical carrier of quantum information.  We examine the bipartite nonlocal correlations of pair of spin-1/2 particle system for the thermal equilibrium states, characterized by local quantum uncertainty (LQU) and local quantum Fisher information (lQFI). The effects of dipolar coupling constants on quantum correlation quantifiers are studied. The DM interaction greatly enhances the quantum correlation in the system whereas the temperature tends to annihilate the amount of quantum correlations.
\end{abstract}
\vspace{1cm}
~~~~~~~~~\noindent{\it Keywords}: Fisher information, Skew information, Quantum correlation, Dipolar system.
\newpage
 \renewcommand{\thefootnote}{*}
\section{Introduction}
Local operations on the composite quantum state can rise nonlocal or global effects which are characteristic features of quantum theory and makes a fundamental departure from classical theory.  Nonlocality is a core inherent reason that quantum information processing (QIP) tasks have great advantages over their classical counterparts. In the early of quantum information theory, entanglement is recognized as the most important manifestation of nonlocality and the only resource for various information processing.  The quantum advantages in deterministic quantum computation with one qubit (DQC1) protocol can be achieved through the separable state (zero entanglement) \cite{Datta2008}. Further, Werner also showed that the entanglement is an incomplete manifestation of nonlocality via Bell inequality \cite{Bell} using a mixed entangled state \cite{Werner1989}. These seminal works  indicate that there  exists  a special kind  of  property in quantum  systems,  which  is  different  from  entanglement  but  still  efficient  for  solving  some  classically  intractable  problems. To understand the complete picture of the nonlocality  of the unentangled (separable) system, Ollivier  and  Zurek  introduced  quantum  discord (QD) as a measure of the quantumness of bipartite state \cite{Ollivier} and QD  plays  a  significant  role  in QIP tasks as well as in the fundamental issues of quantum mechanics. However, QD is quite hard to compute  analytically \cite{Huang}. To overcome this computational issue, an alternate way consists in adopting metric in state space to quantify the minimal distance between a bipartite state and a classically correlated state \cite{Dakic2010,Luo2011}.

Whenever the system is correlated, any local measurements on the multipartite system have a degree of uncertainty which is a primary tool to characterize discord like nonlocal correlation measures for the bipartite quantum system. Based on the local measurements, many nonlocal correlation measures have been introduced such as local quantum uncertainty (LQU) \cite{GirolamiPRL2013}, measurement induced disturbance (MID) \cite{Mista2011}, uncertainty-induced nonlocality (UIN) \cite{Wu2014}, measurement-induced nonlocality (MIN) \cite{Luo2011} and local quantum Fisher information (lQFI) \cite{Kim208}. The skew information is a measure of theoretic-information content in a quantum system, introduced by Wigner and Yanse \cite{Wigner1963}. Further, skew information is also useful in quantification of resources such as quantum correlation \cite{GirolamiPRL2013,Li2016}, asymmetry \cite{Takagi2019,Li2020} and coherence \cite{Yu2015}. Girolami et al. introduced a discord-like correlation measure based on the skew information, namely LQU, which fixes the local ancilla problem with geometric discord \cite{Piani2012} and deeply connected with the Hellinger distance-based correlation measure.  Beyond its importance as a theoretic-information quantifier, LQU is also closely related to quantum Fisher information (QFI) in the context of parameter estimation. Similar to LQU, lQFI is also recognized  as promising candidates to understand the role of nonclassical correlations other than the entanglement in enhancing the precision and efficiency of quantum metrology protocols. Further, the usefulness of lQFI in state classification in the perspective of quantumness is also demonstrated \cite{Kim208} and it is shown that the quantum correlation quantified by the QFI is useful for quantum metrology.
 
 Since, both LQU and QFI are having more practical significance in metrology and the condensed matter systems are regarded as a practical candidate to be used for manipulating QIP tasks. In view of this, it is of great importance to understand the behaviors of LQU and QFI in the physical system, which is useful for manipulating QIP tasks. In this paper, we study the properties of thermal LQU and QFI in a  dipolar spin with the Dzyaloshinsky-Moriya (DM) interaction. The effects of temperature and DM interaction on quantum correlation measures measured by LQU and lQFI are noted. The inequality between the skew information and QFI is also confirmed in the  physical system using LQU and lQFI. 
 
This paper is organized as follows. In Sec. \ref{Sec2}, we review the quantum correlation measures employed here.  In Sec. \ref{Sec3},  we introduce the physical system under our consideration for the present analysis. The role of system parameters, temperature, and DM interaction on quantum correlations  are presented with supporting numerical plots in Sec. \ref{Sec4}.  Finally, in Sec. \ref{concl} we present the conclusion.

\section{Quantum Correlations Quantifiers}\label{Sec2}
In this section, we provide a brief introduction about the quantum correlation quantifier to studied in the two spin-1/2 dipolar system with DM interaction.
\subsection{Local quantum uncertainty}

The local quantum uncertainty (LQU) is a more reliable measure of the quantumness of bipartite state, which goes beyond entanglement. In recent times, the researchers paid wide attention on this discord-like measures. This is essentially due to its  easy computation and the fact that it enjoys all necessary properties of being a faithful measure of quantum correlation. It is shown that LQU is non-zero for the separable state, even in the absence of entanglement. For a bipartite state $\rho$, the LQU is defined as the minimal skew information attainable with a single local measurement. Mathematically, it is defined as \cite{GirolamiPRL2013}, 
\begin{align}
\mathcal{U}(\rho)=~^{\text{min}}_{H^{a}} ~~\mathcal{I}(\rho, H^a\otimes\mathds{1}^b),
\end{align}
where $H^a$ is some local observable on the subsystem $a$, $\mathds{1}^b$ is the $2\times 2$ identity operator acting on the system $b$ and 
\begin{align}
\mathcal{I}(\rho)=-\frac{1}{2}\text{Tr}\left( [\sqrt{\rho}, H^a\otimes\mathds{1}^b]^2 \right)
\end{align}
is the skew information which provides an analytical tool to quantify the information content in the state $\rho$ with respect to the observable $H^a$, $[\cdot, \cdot]$ is the commutator operator.   The  information  content  of $\rho$ about $H^a$ is here  quantified  by  how  much  the  measurement  of $H^a$ on  the  state  is  uncertain. The measurement outcome is certain, only if the state is an eigenvector of $H^a$. On the other hand, if  it  is  a  mixture of eigenvectors of $H^a$,  the uncertainty is only due to imperfect knowledge of the state. For pure bipartite states, the local quantum uncertainty reduces to the linear entropy of entanglement and vanishes for classically correlated states. For $2\times n$ dimensional bipartite system, the closed formula of LQU is computed as 
\begin{align}
\mathcal{U}(\rho)=~ 1- \text{max}\{\omega_1,\omega_2,\omega_3 \}.
\end{align}
Here $\omega_i$ are the eigenvalues of matrix $W$ and whose matrix elements are defined as 
\begin{align}
w_{ij}=\text{Tr}[\sqrt{\rho}(\sigma_i^a\otimes\mathds{1}^b)\sqrt{\rho}(\sigma_i^a\otimes\mathds{1}^b)],~~~~~\text{with}~~~ i,j=1,2,3,
\end{align}
where $\sigma_i$ represents the Pauli spin matrices.
\subsection{Quantum Fisher information}
As a second quantum correlation quantifier, we employ the quantum Fisher information based correlation measure. QFI also quantifies the amount of information contained in state $\rho$ with respect to the observable $H^a$ \cite{Helstrom}. For pure state, the QFI is reduced variance of the observable in the state $\rho$. Further, QFI is the most widely used quantity for characterizing the ultimate accuracy in parameter estimation scenarios. This quantity is also useful in various applications in quantum information theory such as entanglement detection and quantification, quantum metrology, measure of coherence, and identifying phase transition in spin models. In general, for an arbitrary quantum state $\rho_{\theta}$ that depends on the parameter $\theta$, we can define the QFI as
\begin{align}
\mathcal{F}(\rho_{\theta})=\frac{1}{4}\text{Tr}[\rho_{\theta} L_{\theta}^2],  \nonumber
\end{align}
where the symmetric logarithmic derivative $L_{\theta}$ is defined as the solution of the equation
\begin{align}
\frac{d\rho_{\theta}}{d\theta}=\frac{1}{2}( \rho_{\theta} L_{\theta}+L_{\theta}\rho_{\theta} ). \nonumber
\end{align}
The parametric state $\rho_{\theta}$ can be obtained from an initial probe state $\rho$ subjected to a unitary transformation $U = \mathrm{e}^{iH\theta}$ dependent on $\theta$ and generated by a Hermitian operator $H$ i.e., $\rho_{\theta} = U \rho U^{\dagger}$. Following the spectral decomposition $\rho=\sum_{i=1}p_i|\psi_i\rangle \langle \psi_i|$, with $p_i\geq 0$ and $\sum_{i=1}p_i=1$, hence, $ \mathcal{F}(\rho, H )$, is given by \cite{Braunstein,Modi2011}
\begin{align}
\mathcal{F}(\rho, H)=\frac{1}{2}\sum_{i\neq j}\frac{(p_i-p_j)^2}{p_i+p_j}|\langle \psi_i|H |\psi_i\rangle|^2.
\label{QFI}
\end{align}
The quantum correlation measure based on QFI is defined as \cite{Kim208}
\begin{align}
\mathcal{Q}(\rho)=~^{\text{min}}_{H^{a}}~\mathcal{F}(\rho, H),
\end{align}
where $H=H^a\otimes\mathds{1}$. This measure has the desirable properties that any good quantum correlation quantifiers should satisfy. Indeed, it is non-negative, vanishes for zero discord bipartite states (classically correlated states), invariant under any local unitary operation and coincides with the geometric discord for pure quantum states. Choosing the local observable $H^a=\vec{\sigma}.\vec{r}$ with $| \vec{r}|=1 $ and $\vec{\sigma}=(\sigma_x, \sigma_y, \sigma_z)$, the QFI in Eq. (\ref{QFI}) is reformulated as 
\begin{align}
\sum_{i\neq j}\frac{p_ip_j}{p_i+p_j}|\langle \psi_i|H |\psi_i\rangle|^2=\vec{r}^{\dagger}. M. \vec{r},
\end{align}
where the matrix elements of symmetric matrix $M$ are 
\begin{align}
M_{lk}=\sum_{i\neq j}\frac{p_ip_j}{p_i+p_j} \langle \psi_i|\sigma_l\otimes \mathds{1}^b|\psi_j\rangle ~\langle\psi_j|\sigma_k\otimes \mathds{1}^b|\psi_i\rangle, ~~~~\text{with}~~ l,k=x,y,z.
\end{align}
 The closed formula of the quantum correlation measure based on QFI is 
 \begin{align}
  \mathcal{Q}(\rho)=1-\text{max}\{M_1,M_2,M_3 \},
 \end{align}
 where $M_i$ are the eigenvalues of matrix $M$.
\section{The model and thermalization}\label{Sec3}
To understand the behaviors of bipartite quantum correlation, we consider the spin-1/2 particles coupled with dipolar \cite{Reis} and DM interactions \cite{Dzyaloshinskii,Moriya}. The dipolar interaction arises due to the magnetic moment of a spin on another spin located at the nearest site and the DM interaction due to spin-orbit coupling. The Hamiltonian of the system is given as \cite{Reis}
\begin{align}
H=-\frac{1}{3}\vec{S_1} \cdot \overleftrightarrow{\mathcal{D}}\cdot \vec{S_2}+ \vec{D}.(\vec{S_1}\times \vec{S_2}),
\label{ham}
\end{align}
where $\overleftrightarrow{\mathcal{D}}=\text{diag}(\Delta-3\epsilon, \Delta+3\epsilon,-2\Delta )$ is a traceless diagonal tensor, $\Delta$ and $\epsilon$ are the dipolar coupling constants between the spins and $\vec{D}$ is the DM coupling vector. Here we choose the DM coupling along z-axis i.e., $\vec{D}=(0,0,D)$. The sign of $\Delta$ indicates the orientation of the spin: if $\Delta< 0$, the spin is in the $x-y$-plane, whereas if $\Delta> 0$, the spin is directed along the z-axis.

In the standard two-qubit computational basis $\{ |00\rangle,|01\rangle,|10\rangle,|11\rangle \} $, the eigenvalues and corresponding eigenvectors of the Hamiltonian (\ref{ham}) are computed as  
\begin{subequations}\lb{eigen}
	\begin{align}
  \lb{eq1} \lambda_{1,4}=  \frac{1}{6}(\Delta \pm 3\epsilon), ~& ~~~~~~~~~~~~~~~~~~~~~~~~~~~~~~~~\vert \varphi_{1,4}\rangle= \frac{1}{\sqrt{2}}\left(\vert 11 \rangle \pm \vert 00 \rangle\right),~~~~~~~~~~~~\\
   \lb{eq2} \lambda_{2,3}= \frac{1}{6}(\Delta \pm \eta), & ~~~~~~~~~~~~~~~~~~~~~~~~~~~~~~~~\vert \varphi_{2,3}\rangle= \frac{1}{\sqrt{2}}\left(\vert 10 \rangle \pm \vert 01 \rangle\right),
\end{align}
\end{subequations}
where $\eta=\sqrt{9D^2+\Delta^2}$. At the thermal equilibrium, the thermal density operator defined as
\beq\lb{partition1}
\varrho(T)=\frac{1}{\mathcal{Z}}\exp{\left(-\beta H\right)}=\frac{1}{\mathcal{Z}}\sum_{i=1}^4 p_i \vert \varphi_{i}\rangle \langle \varphi_{i}\vert,
\eeq
\begin{figure*}[!ht]
\centering\includegraphics[width=0.6\linewidth]{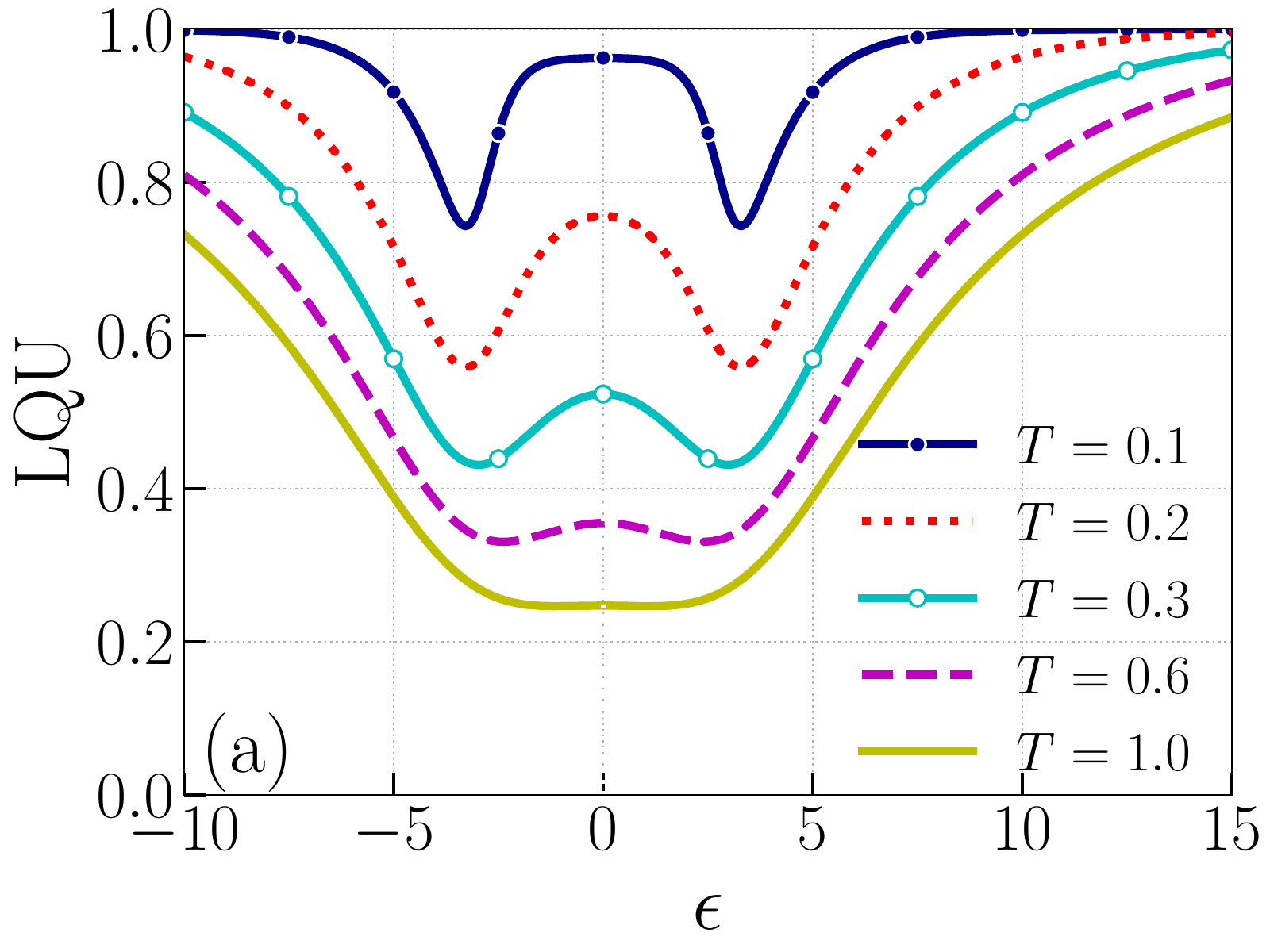}
\centering\includegraphics[width=0.6\linewidth]{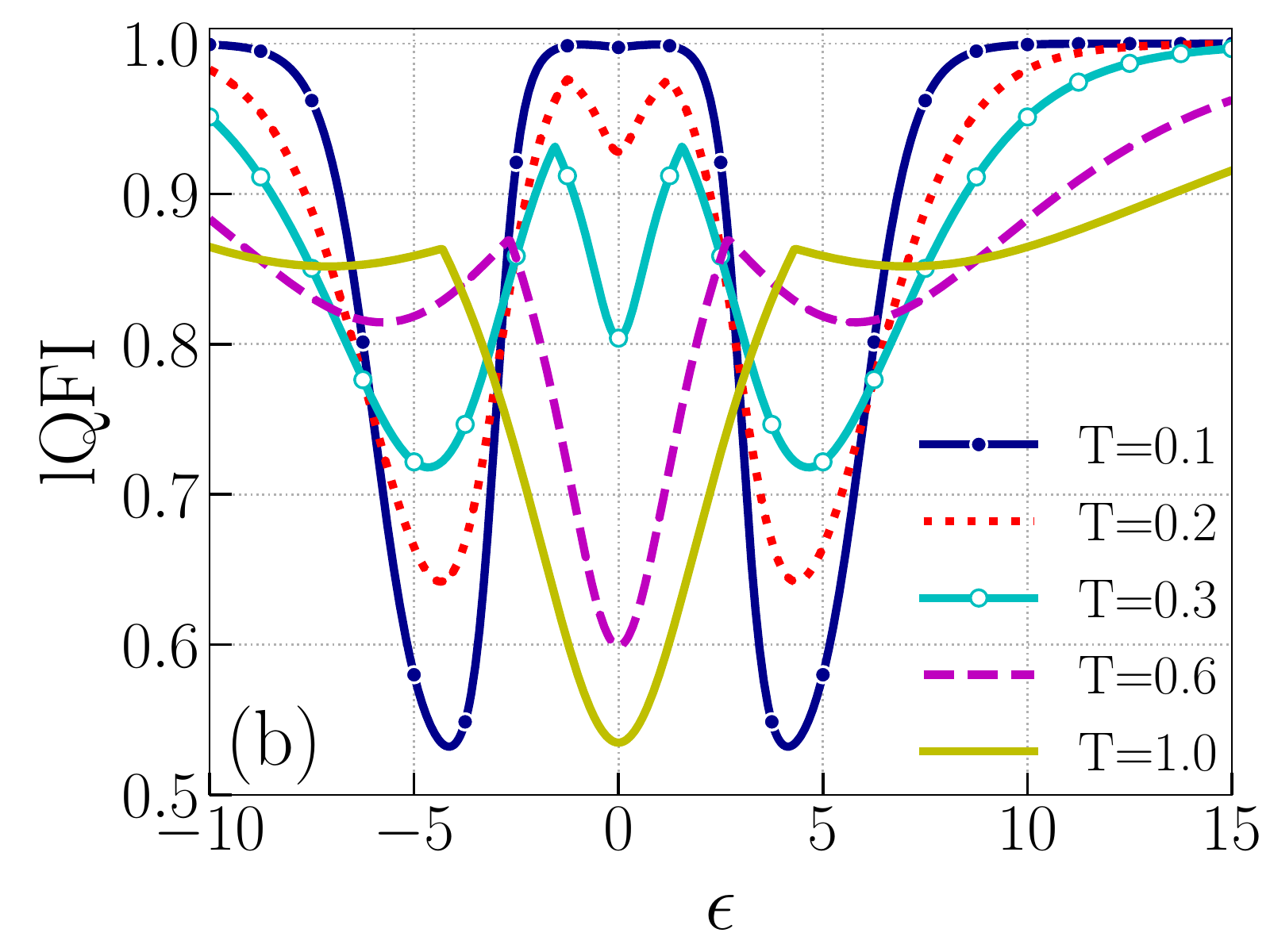}
\caption{(color online) Quantum correlation quantifiers (a) LQU and (b) lQFI as a function of $\epsilon$ for different temperatures.}
\label{fig1}
\end{figure*}
\begin{figure*}[!ht]
\centering\includegraphics[width=0.6\linewidth]{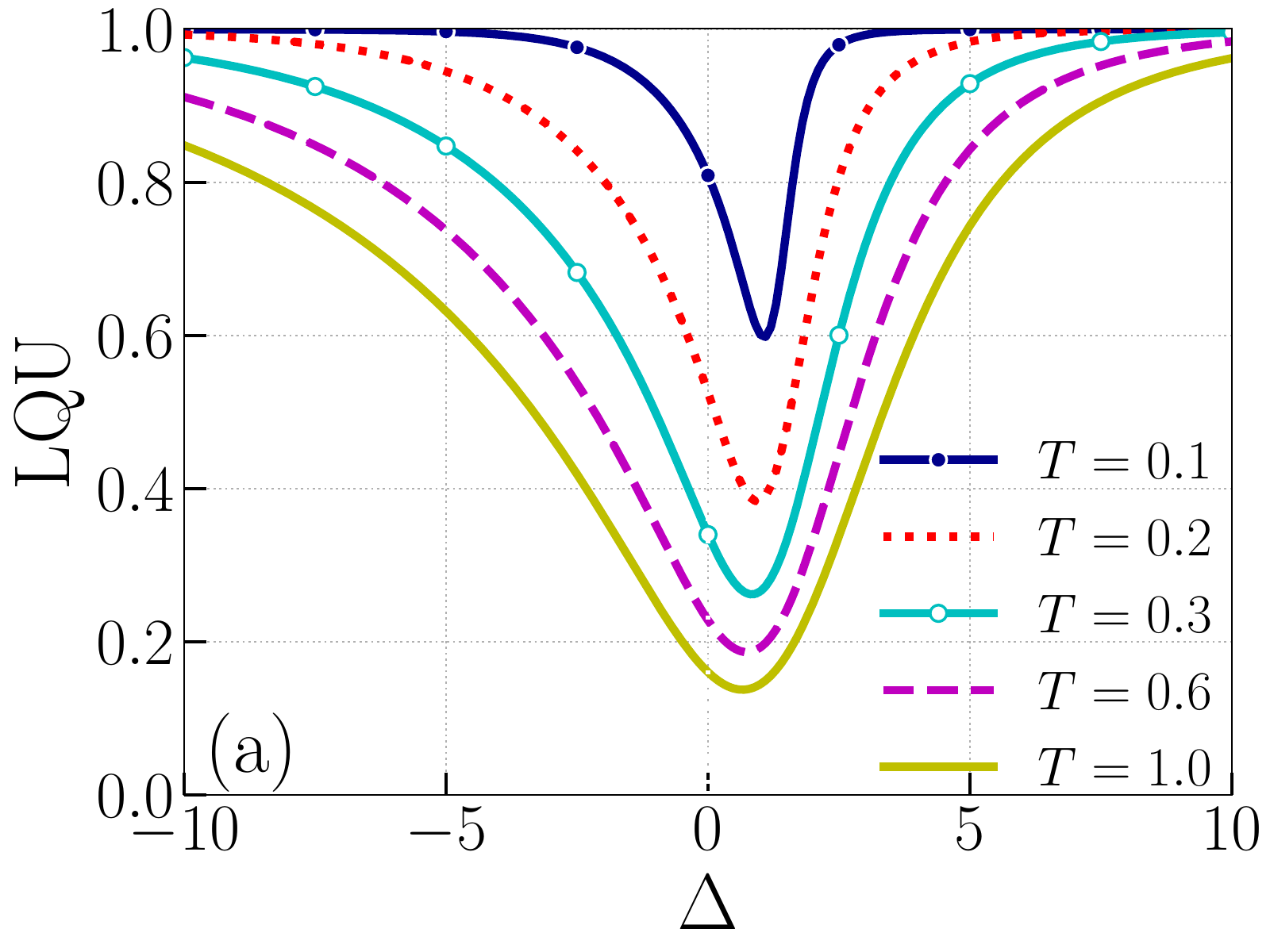}
\centering\includegraphics[width=0.6\linewidth]{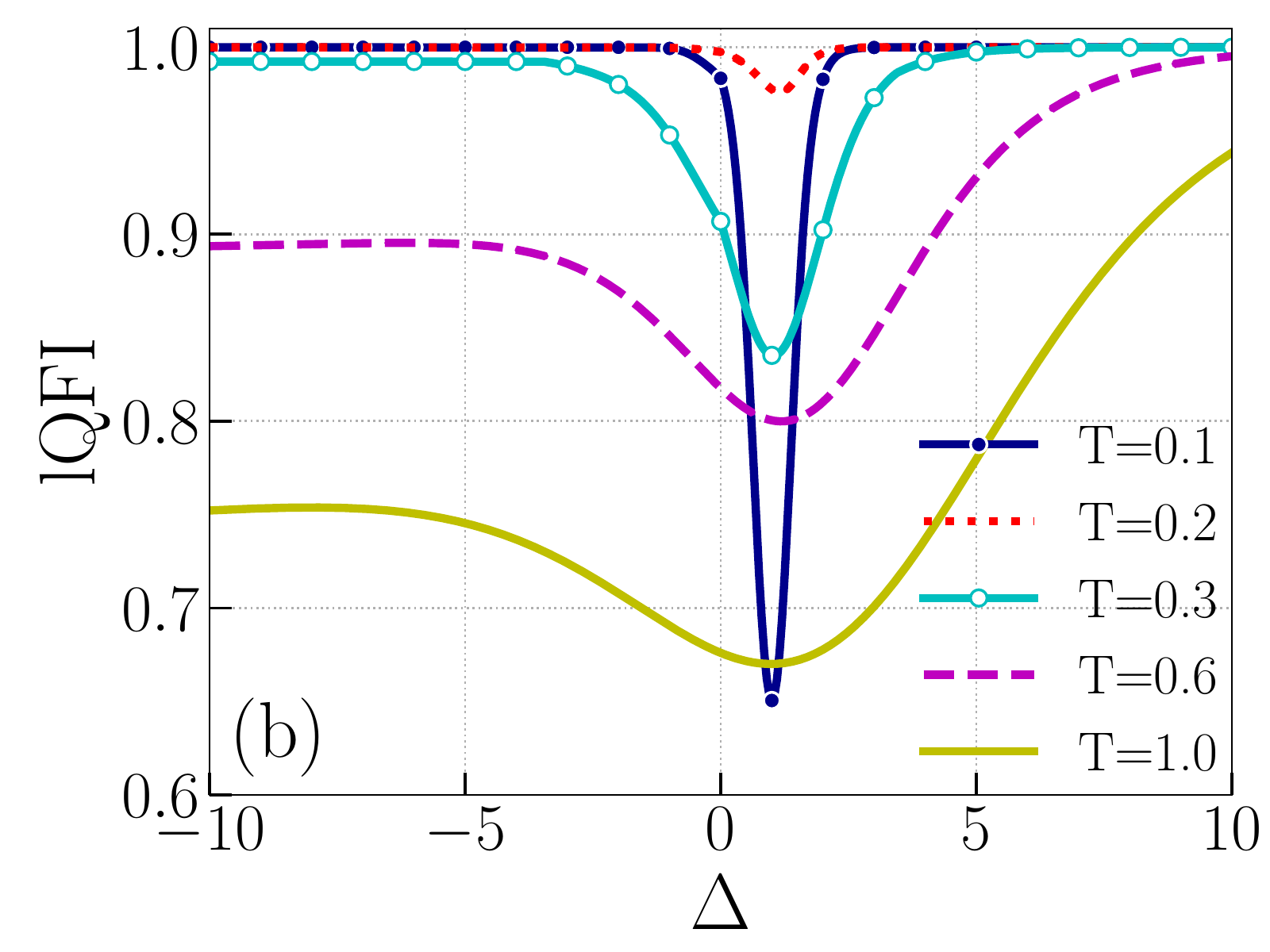}
\caption{(color online) Quantum correlation quantifiers (a) LQU and (b) lQFI as a function of $\Delta$ for different temperatures.}
\label{fig2}
\end{figure*}
where ${\mathcal{Z}} = \text{Tr}\exp\left(-\beta H\right)$ being the canonical ensemble partition function of the system, $p_i$ is eigenvalues of $\varrho(T)$ and the inverse thermodynamic temperature $\beta = 1/(k_BT)$ wherein $k_B$ is the Boltzmann's constant which is considered as unity in the following for simplicity. The thermal density matrix $\varrho(T)$ can be written as
\begin{align}
\varrho(T) = \frac{1}{\mathcal{Z}}\begin{pmatrix}
 \varrho_{11} & 0 & 0 & \varrho_{14} \\
 0 & \varrho_{22} & \varrho_{23} & 0 \\
 0 & \varrho_{23}  & \varrho_{22} & 0 \\
\varrho_{14} & 0 & 0 &  \varrho_{11}
\end{pmatrix},
\label{thermal}
\end{align}
where are the matrix elements 
$\varrho_{11}=\mathrm{e}^{\frac{-\beta \Delta}{6}}\text{cosh}(\frac{\beta \epsilon}{2})$, $\varrho_{14}=-e^{\frac{-\beta \Delta}{6}}\text{sinh}(\frac{\beta \epsilon}{2})$, $\varrho_{22}=\mathrm{e}^{\frac{\beta \Delta}{6}}\text{cosh}(\frac{\beta \eta}{6})$ and $\varrho_{23}=-\mathrm{e}^{\frac{\beta \Delta}{6}}\text{sinh}(\frac{\beta \eta}{6})$. The partition function of the system is $\mathcal{Z}=2\mathrm{e}^{\frac{-\beta \Delta}{6}}\text{cosh}(\frac{\beta \epsilon}{2})+2\mathrm{e}^{\frac{\beta \Delta}{6}}\text{cosh}(\frac{\beta \eta}{6})$. It is worth mentioning that $\varrho(T)$ commute with $H$ and they can have a common set of eigenfunctions in Eq. (\ref{eigen}). The eigenvalues of $\varrho(T)$ are computed as 
\begin{align}
p_{1,4}=\frac{\varrho_{11}\pm \varrho_{14}}{\mathcal{Z}} ~~~ \text{and} ~~~ p_{2,3}=\frac{\varrho_{22}\pm \varrho_{23}}{\mathcal{Z}}.
\end{align}
\section{Results and Discussions}\label{Sec4}
In what follows, we study the behaviors of quantum correlation characterized by local quantum uncertainty (LQU) and  local quantum Fisher information (lQFI). To understand the effects of dipolar coupling constants on quantum correlations, we plot the correlation measures in figure \ref{fig1} as a function of $\epsilon$ for various values of the temperatures in the absence of DM interaction. It is observed that LQU is an even function of $\epsilon$.  It is clearly seen that for fixed values of temperature, LQU decreases with $|\epsilon|$ and  after reaching a minimal value LQU increases monotonically, and reaches the maximum 1 for sufficiently larger $|\epsilon|$. We observe that increasing temperature values tend to reduce the amount of quantum correlations in the system. On the other hand, the companion quantity lQFI is plotted in figure \ref{fig1}(b), we observed that $\mathcal{Q}(\varrho)$  is also even function of $\epsilon$. For $T=0.1$, QFI is maximum in a small region of $|\epsilon|$, and reaches a minimal value at $\epsilon\approx \pm4$. Due to further increment of $|\epsilon|$,  lQFI increases monotonically and reaches a maximum for sufficiently larger values of $|\epsilon|$. Here also we noted that increasing the values of temperature the amount of correlation captured by lQFI decreases.  It is worth mentioning that QFI is bounded by the skew information. Further, we observe that lQFI is always greater than LQU for all the parametric regions. Our results also confirm the inequality  $\mathcal{I}(\rho,H) \leq \mathcal{F}(\rho, H) \leq 2\mathcal{I}(\rho, H)$ and resulting from that $\mathcal{U}(\rho)\leq\mathcal{Q}(\rho)$.

Next, we analyze the effect of $\Delta$ on the quantum correlation measures for the fixed values of temperature. For this we set $\epsilon=2~ \text{and}~ D=0$, it is observed that both the correlation measures LQU and lQFI show qualitatively similar behavior which is shown in figure \ref{fig2}. When the spins are oriented in the \textit{x-y} plane, the quantum correlations are almost constant, whereas both the measures are sensitive to the system parameter while the spins oriented along the \textit{z}-direction. While increasing $\Delta$, both the measures decreases to a minimal value, then increases with $\Delta$ and saturates at  maximal values for higher values of $\Delta$. Here also we observe that the temperature reduces the quantum correlation between the subsystems.
\begin{figure*}[!ht]
\centering\includegraphics[width=0.6\linewidth]{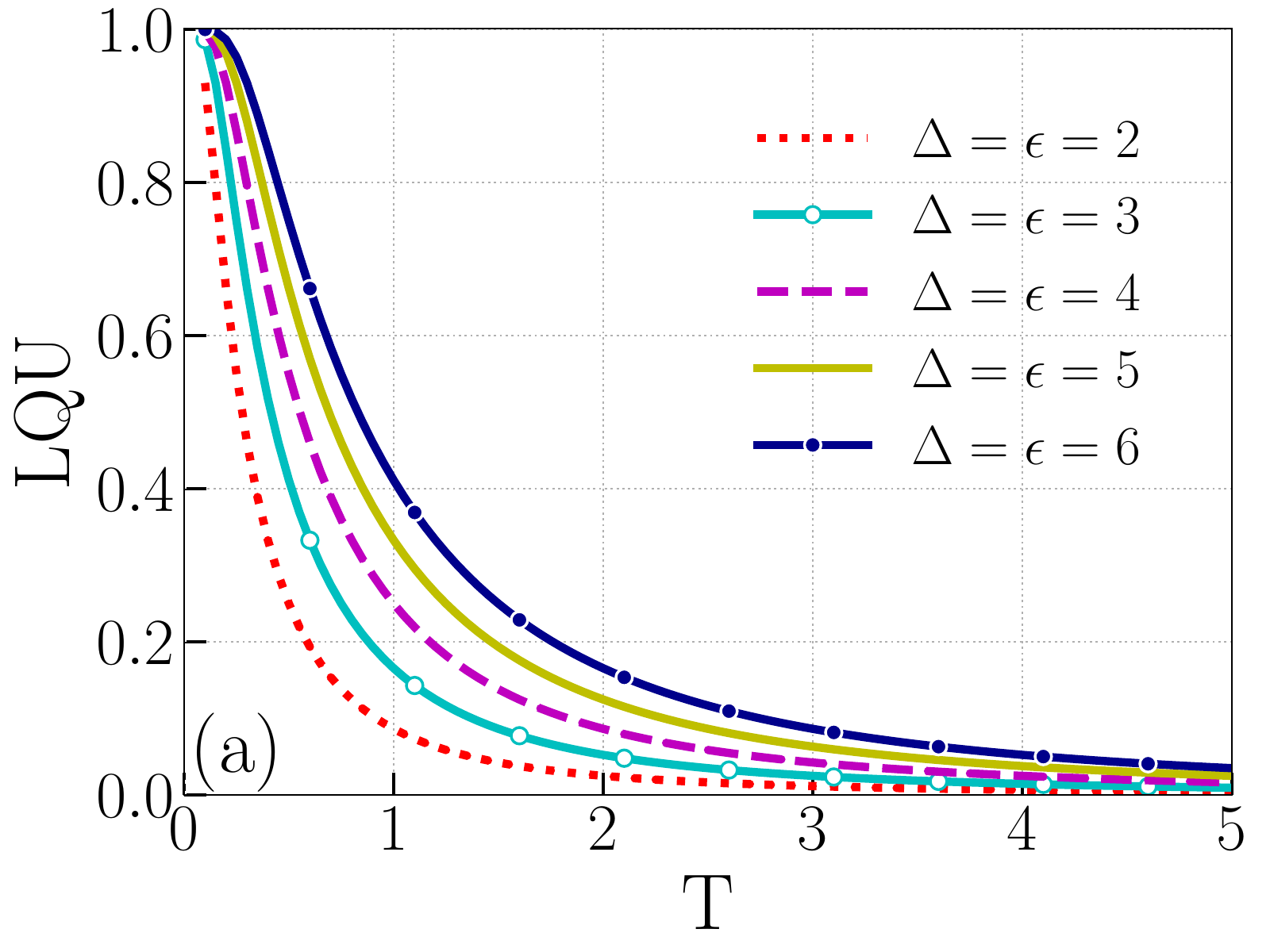}
\centering\includegraphics[width=0.6\linewidth]{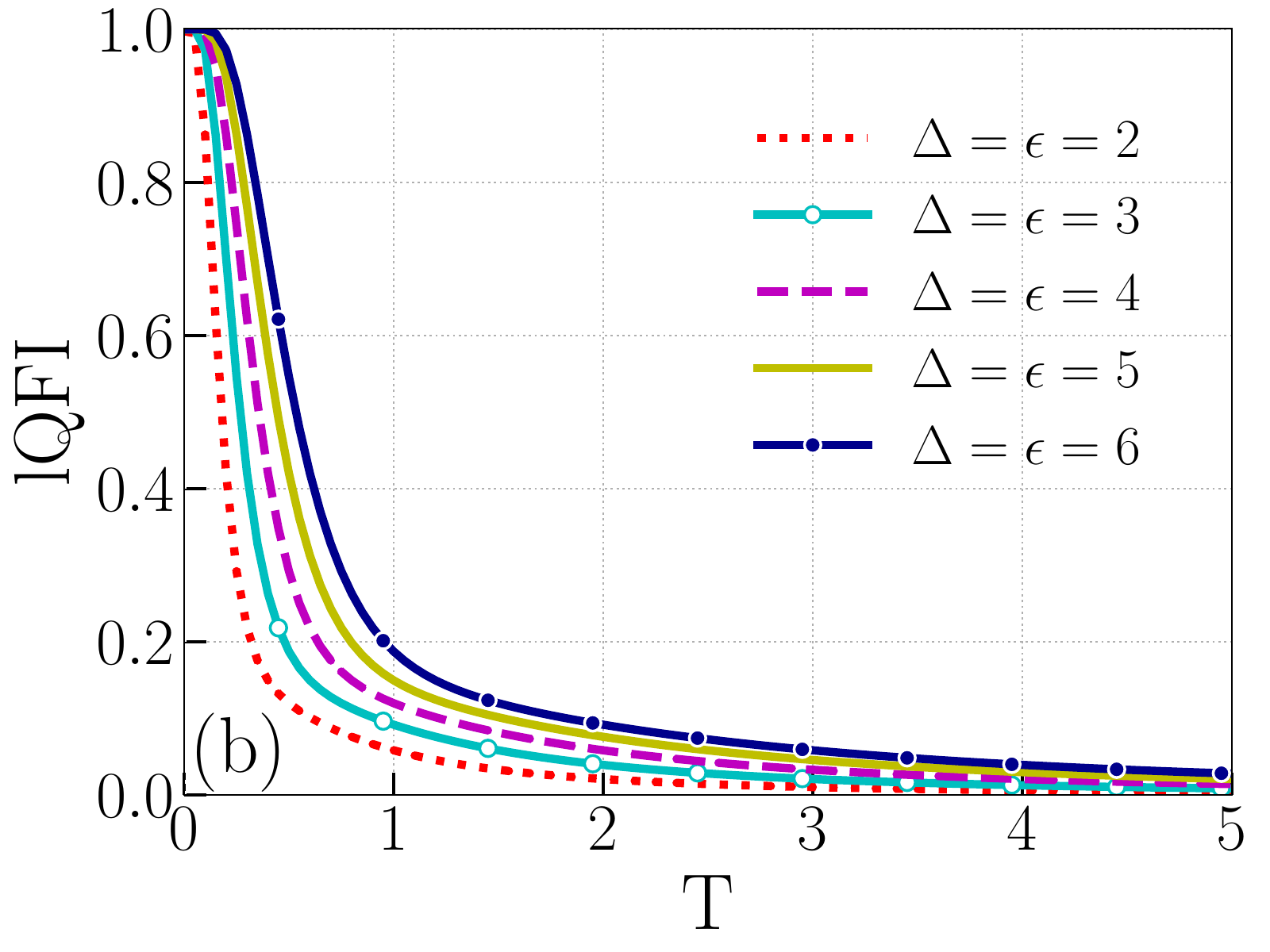}
\caption{(color online) Temperature dependence of quantum correlation measures for different dipolar coupling constants $\epsilon$ and $\Delta$. }
\label{fig3}
\end{figure*}

Let us now analyze the effects of temperature on LQU and lQFI for the different values of $\Delta~ \text{and}~ \epsilon$ in the absence of DM interaction which is shown in figure \ref{fig3}. At $T=0$, the  ground state energy is 0, and the corresponding ground state $\vert \varphi_{3}\rangle$ is a maximally entangled state. It is observed that both measures are maximum at $T=0$, implying that the maximally entangled state corresponds to maximally correlated. As temperature increases, $\mathcal{U}(\varrho)$ decreases monotonically with temperature and vanishes at a higher temperature. Another correlation measure  $\mathcal{Q}(\varrho)$ also shows similar  monotonically decreasing behavior with an increase of temperature. At higher temperatures, the quantum correlation captured by lQFI drops to zero. Increasing the values of dipolar coupling constants $\Delta~ \text{and}~ \epsilon$, both the measure increases for fixed temperature.

\begin{figure*}[!ht]
\centering\includegraphics[width=0.6\linewidth]{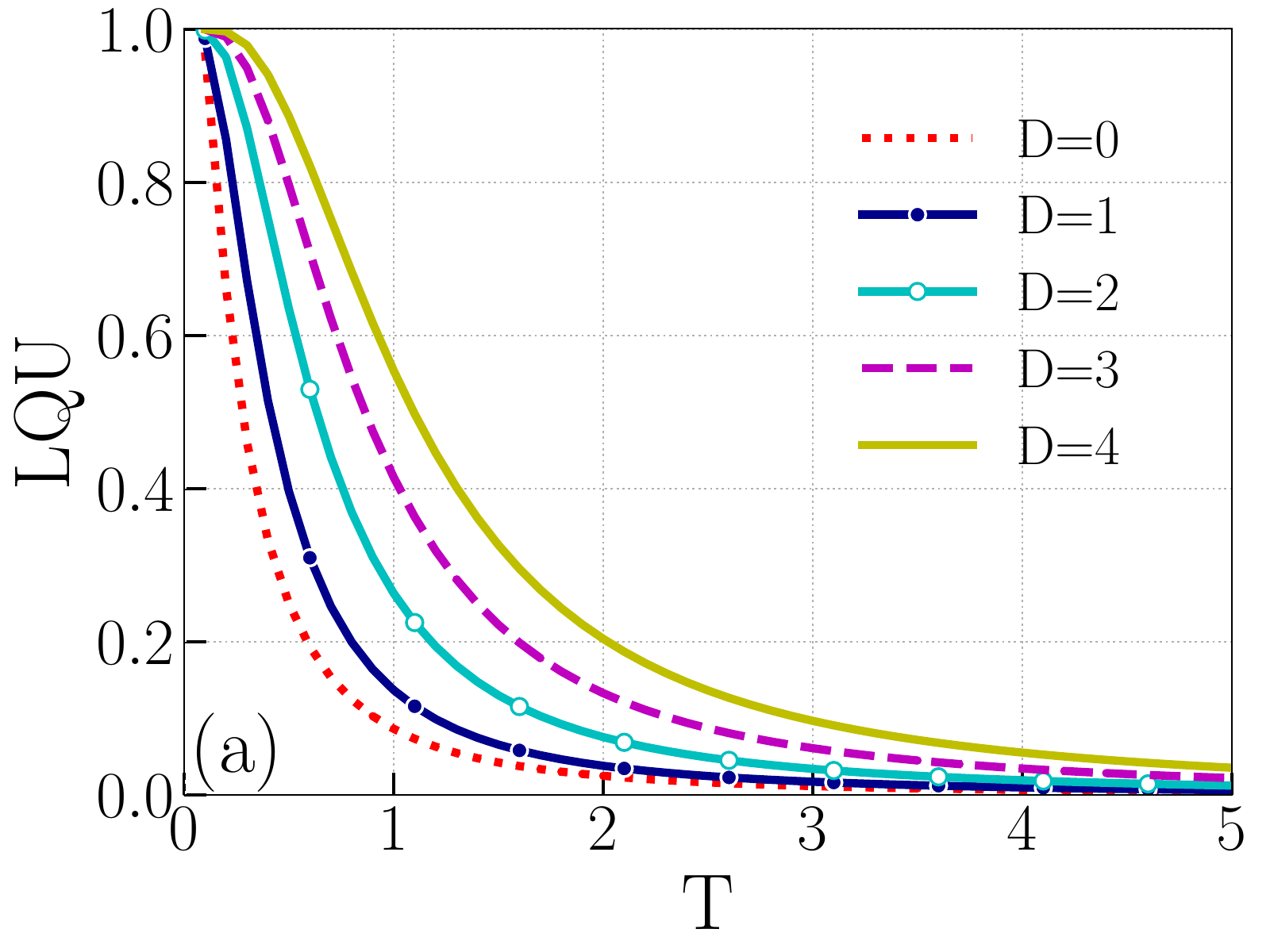}
\centering\includegraphics[width=0.6\linewidth]{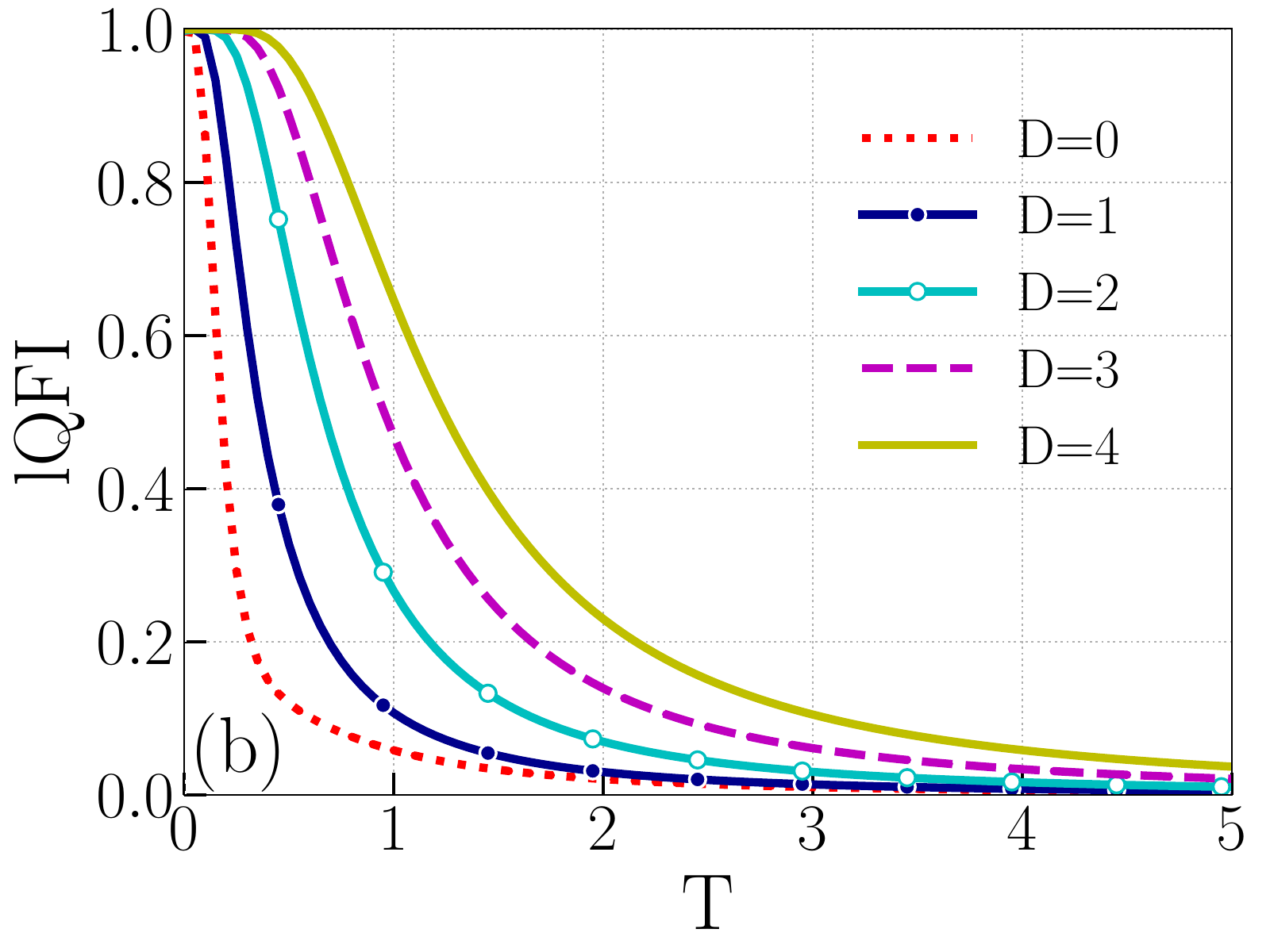}
\caption{(color online) Temperature dependence of quantum correlation measures for different DM interaction parameters $D$. }
\label{fig4}
\end{figure*}

\begin{figure*}[!ht]
\centering\includegraphics[width=0.6\linewidth]{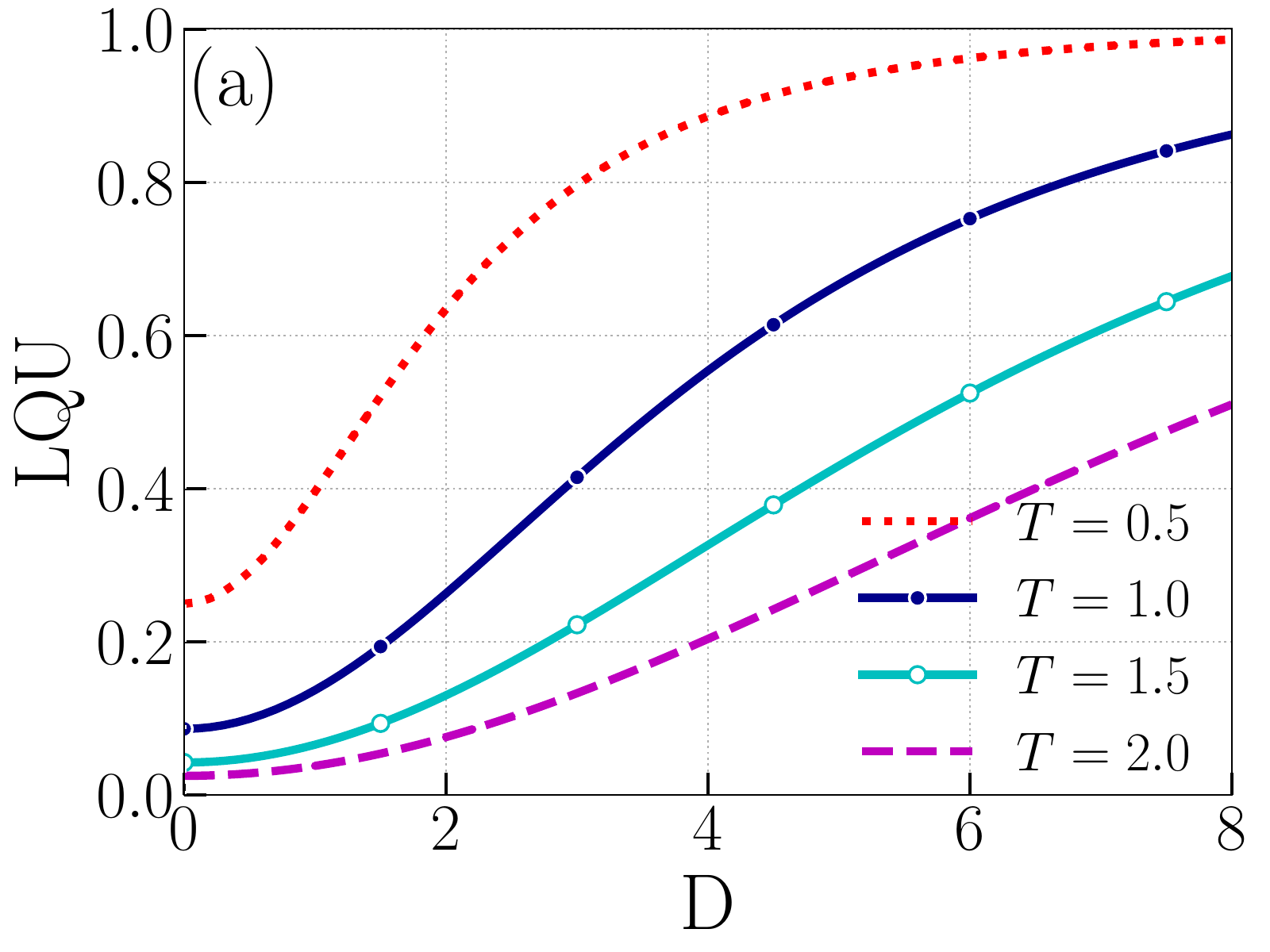}
\centering\includegraphics[width=0.6\linewidth]{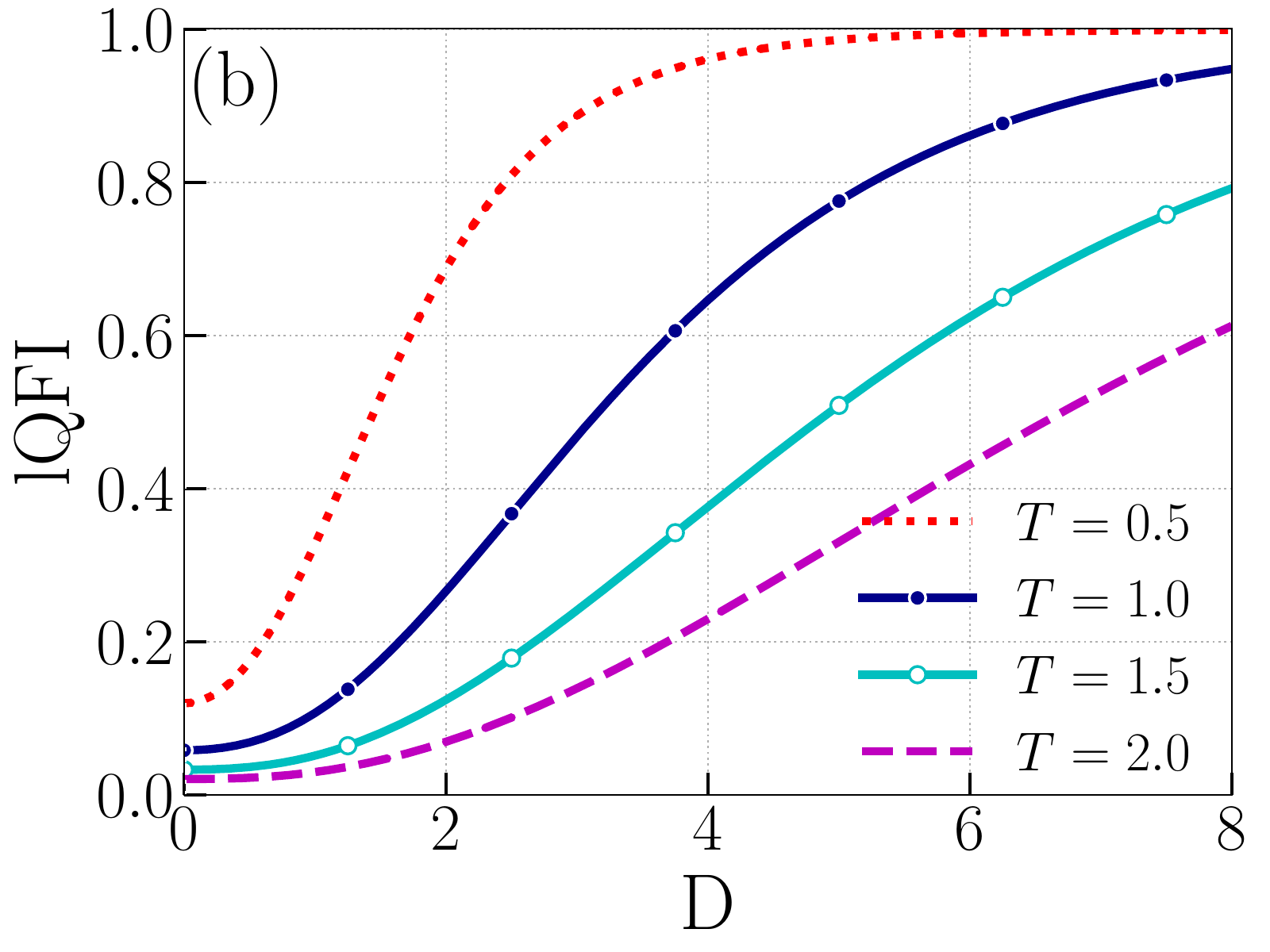}
\caption{(color online) Quantum correlation measures (a) LQU and (b)  lQFI as  a function of DM interaction for different temperatures $T$.}
\label{fig5}
\end{figure*}

In order to see the effect of DM interaction on quantum correlation quantifiers LQU and lQFI, we set $\Delta=\epsilon=2$ and plot the correlation measures as a function of temperature for different values of DM in interaction in figure \ref{fig4} . Here also we observe the monotonically decreasing behavior of both the measures with the temperature. It is clearly seen that the quantum correlation in the system increases with DM interaction. To enhance our understanding on the effects of DM interaction $D$, we plot the correlation measures as a function of $D$ for different temperature values and fixed parameters $\Delta=\epsilon=2$. In general, the DM interaction is a week interaction,  but in order to explore the effect of DM interaction on the quantum correlation at different temperatures, we take the value up to 8. From figure \ref{fig5}, it is found that for different temperatures, as $D$ increases the quantum correlation increases monotonically and reaches the maximum  when DM interaction $D$ is large enough. With the temperature increasing, the maximum of LQU  and lQFI decreases at $D=0$. A similar analysis is done for entanglement in the considered physical system, it is observed that DM interaction decreases the entanglement between the spins \cite{Habiballah}, whereas the correlation measures beyond entanglement (LQU and lQFI) strengthen by the DM interaction.  From the above observation, we observe that the DM interaction can act as a catalyst to enhance the quantum correlation in the system.

\section{Conclusion}
\label{concl}

To summarize, we have studied the quantum correlation measured by local quantum uncertainty and quantum Fisher information in a two spin-1/2 dipolar system in  presence  of Dzyaloshinski–Moriya  interaction  induced  by  the  spin–orbit coupling in thermal equilibrium. A  comparative  study  is  achieved  to  analyze  the  quantum  correlations  presented  in  our  model considered. The effects of dipolar coupling constants on quantum correlation quantifiers are highlighted. We have shown that the DM interaction enhances the quantum correlation in the system and thermal effects tend to decrease the quantum correlation. 

Further, our results claim that the addition of DM interaction in the physical system can improve the efficiency of parameter estimation. 
\noindent

\section*{Acknowledgment}
This work has been financially supported by the Council of Scientific and Industrial Research (CSIR), Government of India, under Grant No. 03(1444)/18/EMR-II.

\end{document}